# Sensitivity of entangled photon holes to loss and amplification

J.D. Franson

*Physics Department, University of Maryland, Baltimore County, Baltimore, MD 21250*

Energy-time entangled photon holes are shown to be relatively insensitive to photon loss due to absorption by atoms whose coherence times are longer than the time delays typically employed in nonlocal interferometry (a fraction of a nanosecond). Roughly speaking, the excited atoms do not retain any significant "which-path" information regarding the time at which a photon was absorbed. High-intensity entangled photon holes can also be amplified under similar conditions. Decoherence does occur from losses at beam splitters, and these results show that photon loss cannot always be adequately modeled using a sequence of beam splitters. These properties of entangled photon holes may be useful in quantum communications systems where the range of the system is limited by photon loss.

## I. INTRODUCTION

Entangled photon holes [1-4] are a new form of entanglement in which the absence of photons in two separated beams is correlated in a nonclassical way. Photon holes can be entangled in energy and time [1], which allows them to violate Bell's inequality in two distant interferometers [5-14]. Since the photon holes correspond to the absence of photons, one might naively suspect that they may be less sensitive to photon loss than are pairs of entangled photons. It is shown here that high-intensity entangled photon holes are relatively insensitive to photon loss if an absorbing medium does not retain any significant "which path" information regarding the time at which a photon was absorbed. Entangled photon holes can also be amplified under similar conditions aside from the effects of spontaneous emission noise.

The concept of entangled photon holes can best be understood by analogy with the generation of entangled pairs of photons, as illustrated in Fig. 1a. Here a nonlinear crystal has a small probability of annihilating a single photon from a pump laser beam and creating a pair of photons. The photons are created at essentially the same time, but there is a coherent superposition of times at which they may have been created, which corresponds to an energy-time entangled state. In this case the two output beams are initially "empty" and the added photons are entangled with each other.

Entangled photon holes [1] can be viewed as the negative image of down-conversion, as illustrated in Fig. 1b. Here two laser beams pass through a medium that can absorb two photons but not one. Pairs of photons are absorbed at essentially the same time, creating a pair of holes in the output beams which originally corresponded to uniform probability amplitudes. There is a coherent superposition of the times at which the pair of photons may have been removed to create a pair of holes, which also corresponds to an energy-time entangled state.

Our first experiments [2] on entangled photon holes used weak laser beams that contained less than one photon on average, while subsequent experiments have involved up to five photons [4]. In this paper, we will consider coherent states that contain moderate numbers of photons ($\sim 10^3$, for example) but are still relatively weak compared to most classical experiments. It is well known that coherent states can be attenuated by an absorptive medium or amplified in various ways with only a relatively small loss of coherence, and this forms the intuitive basis for the analysis described below. The amount of decoherence depends on how much information is left in the medium regarding the time at which an atomic transition occurred.

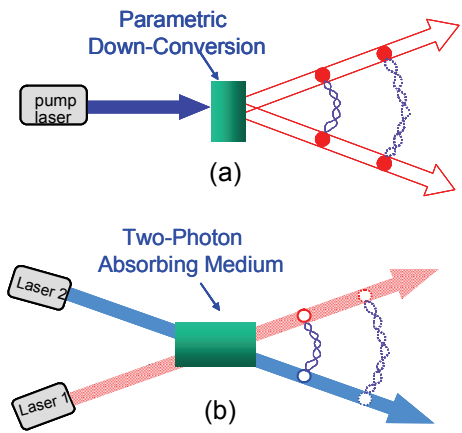

Fig. 1. (a) Generation of entangled photon pairs using parametric down-conversion. (b) Generation of entangled photon holes using two-photon absorption.

Entangled photon holes states of this kind are described in more detail in Section II. The effects of photon absorption in an atomic medium are considered in Section III, where it is shown that entangled photon holes do not undergo any decoherence in the limit of long atomic lifetimes. Section IV obtains similar results for the amplification of entangled photon holes using an inverted atomic medium. Section V includes the effects of decoherence of the excited atomic states and provides estimates of the acceptable photon losses under those conditions. Section VI provides a summary and conclusions.

## II. ENTANGLED STATE OF INTEREST

There may be a variety of entangled photon hole states with similar properties. Here we will consider what is probably the most straightforward example in which the entangled photon holes exist in a uniform background of coherent states. This simplifies the analysis, while the properties of this state should have much in common with other possible forms of entangled photon holes.

Photon holes consist of the correlated absence of photons in an otherwise constant background of probability amplitudes. Although we are primarily interested in the holes, they can only be described by first considering the background in which they reside. In order to do that, we will consider an operator $\hat{A}^\dagger(x_1)$ that creates a single photon in one of the two beams (beam 1) in the form of a Gaussian packet centered at $x = x_1$ as illustrated in Fig. 2:

$$\hat{A}^\dagger(x_1) \equiv \sum_k c_k(x_1)\hat{a}_k^\dagger. \tag{1}$$

Here $x$ is the direction of propagation and the operator $\hat{a}_k^\dagger$ creates a plane-wave photon with wave vector $k$ along the x axis in beam 1.

The central frequency of the Gaussian packet is chosen to be $\omega_0$ and it will be assumed that the Fourier coefficients $c_k(x_1)$ give

$$\sum_k c_k(x_1)e^{ikx} = c_0 e^{ik_0 x} e^{-(x-x_1)^2/2\sigma_p^2}. \tag{2}$$

Here $k_0 = \omega_0/c$, $\sigma_p$ is the width of the Gaussian, and $c_0$ is a suitable normalization constant [15]. Acting with this operator on the vacuum state $|0\rangle$ gives the corresponding single-photon state

$$|\psi_p(x_1)\rangle = \sum c_k(x_1)\hat{a}_k^\dagger|0\rangle. \tag{3}$$

We will also define a similar operator $\hat{A}^\dagger(x_2)$ that creates a single photon in a Gaussian packet in beam 2.

Now consider a coherent state [16] $|\alpha(x_1)\rangle$ with a large number of photons in the Gaussian mode described above:

$$|\alpha(x_1)\rangle = e^{-\alpha^*\alpha/2} e^{\alpha \hat{A}^\dagger(x_1)}|0\rangle. \tag{4}$$

Here $\alpha$ is a complex number that determines the amplitude of the coherent state. This can be rewritten in the form

$$|\alpha(x_1)\rangle = e^{-\alpha^*\alpha/2} e^{\alpha \sum c_k(x_1)\hat{a}_k^\dagger}|0\rangle = e^{-\alpha^*\alpha/2} \prod_k e^{\alpha c_k(x_1)\hat{a}_k^\dagger}|0\rangle$$
$$= \prod_k |\alpha c_k(x_1)\rangle. \tag{5}$$

Here $|\alpha c_k(x_1)\rangle$ is a single-mode coherent state in beam 1 with wave vector $k$ and amplitude $\alpha_k = \alpha c_k(x_1)$. We define a coherent state $|\alpha(x_2)\rangle$ in beam 2 in a similar way.

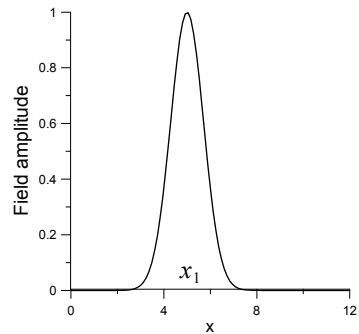

Fig. 2. Probability amplitude to detect a single photon generated by the operator $\hat{A}^\dagger(x_1)$ as a function of position in beam 1 (arbitrary units).

The operators $\hat{A}^\dagger(x_1)$ and $\hat{A}^\dagger(x_2)$ can now be used to define an entangled photon hole state given by

$$|\psi_F\rangle = \chi \iint dx_1 dx_2 f(x_1, x_2)|\alpha(x_1)\rangle|\alpha(x_2)\rangle$$
$$= \chi \iint dx_1 dx_2 f(x_1, x_2) \prod_k |\alpha c_k(x_1)\rangle \prod_p |\alpha c_p(x_2)\rangle. \tag{6}$$

Here $p$ denotes the wave vector in beam 2, the subscript $F$ refers to the fact that this is the state of the field, and $\chi$ is a normalizing constant. The function $f(x_1, x_2)$ can be taken to be any smooth function with the property that

$$f(x_1, x_2) = 0 \quad \text{iff} \quad |x_1 - x_2| < d. \tag{7}$$

The parameter $d$ in Eq. (7) is assumed to be much larger than the width $\sigma_p$ of the Gaussian packets. This ensures that the probability amplitude to detect a photon in each beam at the same relative location ($x_1 = x_2$) will be exponentially small, which is responsible for creating the

"holes" in the field. For simplicity, it will be assumed that $f(x_1, x_2) = 1$ for $|x_1 - x_2| \gg d$.

The properties of the entangled photon hole state of Eq. (6) are illustrated in Fig. 3. The probability of detecting one or more photons simultaneously in both beams is plotted as a function of the difference $\Delta x$ in the distances from the source in beams 1 and 2. The fact that $f(x_1, x_2) = 0$ for $\Delta x = 0$ ensures that no photons will be detected at the same distance from the source. This is similar to the properties of the weak photon holes considered in Refs. [1] and [2], except that here the uniform background in which the holes reside consists of a constant probability amplitude for coherent states centered at locations $x_1$ or $x_2$ in the two beams. Since the photons propagate at the speed of light, we can also think of them in the time domain as not being present at the same time if the distances are equal.

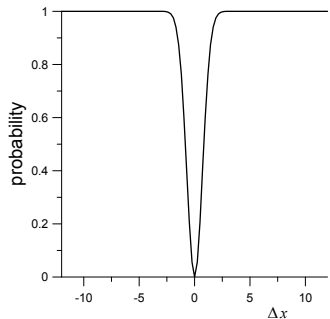

Fig. 3. A plot of the probability of simultaneously detecting one or more photons in both beams 1 and 2 as a function of the difference $\Delta x$ in the distance from the source. The joint detection probability is zero for $\Delta x = 0$, while there is a uniform probability amplitude to detect one or more photons in each beam for $\Delta x \gg d$. (Arbitrary units.)

The entangled state of Eq. (6) can be viewed as a generalization of the energy-time entangled states used in the two-photon interferometer that I previously proposed [5]. The main differences are the form of the function $f(x_1, x_2)$ and the nature of the creation operators (single photons or coherent states). Eq. (6) is a form of entangled Schrodinger cat state [17, 18]. States similar to Eq. (6) can be generated using the Kerr effect and displacement operations, as will be discussed in a subsequent paper. Entangled photon hole states with a periodic (pulsed) background analogous to the experiments of Ref. [2] can also be generated.

### III. ABSORPTION BY IDEAL ATOMS

In most quantum communications applications, the entangled photon holes will have to propagate some distance through an optical fiber or other medium, such as the atmosphere in free-space quantum key distribution. It will be found that the amount of decoherence that occurs due to photon loss will be strongly dependent on the nature of the medium. In this section, the effects of loss due to the absorption of photons in an atomic medium will be considered. It will be found that no decoherence occurs in the idealized case in which there is no significant decay or dephasing of the excited atoms over the time interval of interest, which is typically a fraction of a nanosecond. As a result, the entangled photon holes can be viewed as an example of a decoherence-free subspace [19-21] under the appropriate conditions. The effects of atomic decay and decoherence will be considered in Section V.

The entangled state of Eq. (6) will be taken to be the initial state of the field before it is incident on the absorbing medium, which is assumed to consist of a large number of two-level atoms as illustrated in Fig. 4a. The interaction with any individual atom will be characterized by a matrix element $\sim \varepsilon$, where $\varepsilon$ is assumed to be sufficiently small to allow the use of perturbation theory, for example.

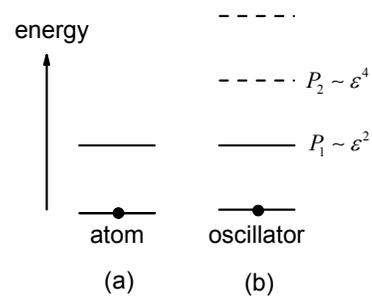

Fig. 4. (a) Two-level atoms that can absorb photons from entangled photon hole states. (b) Harmonic oscillators used to represent the atoms when the coupling is sufficiently small that the probability $P_2$ to occupy the second excited state is negligibly small.

In the limit of small $\varepsilon$, the same results would be obtained if the atoms were replaced by harmonic oscillators as indicated in Fig. 4b. The probability amplitude that any oscillator will be excited to its second excited state is $\sim \varepsilon^2$ while the corresponding probability is $\sim \varepsilon^4$ and negligible. Alternatively, we could have simply assumed that the "environment" that decoheres the entangled photon holes actually consists of a set of harmonic oscillators, as is commonly done. The harmonic oscillators will be chosen to have the same frequency $\omega_0$ as the central frequency of the Gaussian packets that describe the field.

The harmonic oscillators will be labeled with indices $i$ in beam 1 and $j$ in beam 2. All of the harmonic oscillators will be assumed to be in their ground state initially, which corresponds to coherent states $|\beta_i\rangle$ and $|\beta_j\rangle$ with initial amplitudes $\beta_i = \beta_j = 0$ for all $i$ and $j$.

Thus the initial state $|\psi_0\rangle$ of the entire system is a superposition of products of coherent states given by

$$|\psi_0\rangle = \chi \iint dx_1 dx_2 f(x_1, x_2) \prod_k |\alpha c_k(x_1)\rangle \prod_p |\alpha c_p(x_2)\rangle \times \prod_i |\beta_i\rangle \prod_j |\beta_j\rangle. \quad (8)$$

In the usual rotating wave approximation, an atom or harmonic oscillator can absorb a photon and make a transition from its ground state to its first excited state. The corresponding Hamiltonian $\hat{H}$ in the dipole approximation has the form

$$\begin{aligned}\hat{H} &= \sum_k \hbar \omega_k \hat{a}_k^\dagger \hat{a}_k + \sum_p \hbar \omega_p \hat{a}_p^\dagger \hat{a}_p + \sum_i \hbar \omega_0 \hat{b}_i^\dagger \hat{b}_i \\ &+ \sum_j \hbar \omega_0 \hat{b}_j^\dagger \hat{b}_j + \sum_{k,i} \left( h_{ki} e^{-ikx_i} \hat{a}_k^\dagger \hat{b}_i + h_{ki}^* e^{ikx_i} \hat{b}_i^\dagger \hat{a}_k \right) \\ &+ \sum_{p,j} \left( h_{pj} e^{-ipx_j} \hat{a}_p^\dagger \hat{b}_j + h_{pj}^* e^{ipx_j} \hat{b}_j^\dagger \hat{a}_p \right).\end{aligned} \quad (9)$$

The zero-point energies have no effect here and have been omitted. The operators $\hat{b}_i^\dagger$ and $\hat{b}_i$ represent the usual raising and lowering operators for the harmonic oscillators in beam 1 with a similar definition for beam 2, while $\omega_k = ck$ and $\omega_p = cp$ are the angular frequencies of the photons. The location of atom $i$ in beam 1 has been denoted by $x_i$ while the location of atom j in beam 2 has been denoted by $x_j$. The constants $h_{ki}$ and $h_{pj}$ are the matrix elements of the Hamiltonian for the corresponding transitions. For simplicity, we will assume that all the matrix elements have the same value over the range of frequencies contained in the photon wave packets, which is usually a good approximation.

We can now make use of a theorem by Glauber [22], who showed that a product of coherent states evolving in time in accordance with the Hamiltonian of Eq. (9) will remain a product of coherent states at all subsequent times. The amplitudes of the coherent states become time dependent and they will be denoted as follows in order to simplify the notation:

$$\begin{aligned}\alpha c_k(x_1) &\to \alpha c_k(x_1, t) \equiv \alpha_k'(x_1) \\ \alpha c_p(x_2) &\to \alpha c_p(x_2, t) \equiv \alpha_p'(x_2) \\ \beta_i &\to \beta_i(x_1, t) \equiv \beta_i'(x_1) \\ \beta_j &\to \beta_j(x_2, t) \equiv \beta_j'(x_2).\end{aligned} \quad (10)$$

Here we have combined the product of $\alpha$ and $c_k(x_1, t)$ into a single variable $\alpha_k'(x_1)$. The notation $\beta_i'(x_1)$ and $\beta_j'(x_2)$ has been used to indicate that these amplitudes may in general depend on the value of $x_1$ and $x_2$. The time-dependent state of the system is then given by

$$|\psi(t)\rangle = \chi \iint dx_1 dx_2 f(x_1, x_2) \prod_k |\alpha_k'(x_1)\rangle \prod_p |\alpha_p'(x_2)\rangle \times \prod_i |\beta_i'(x_1)\rangle \prod_j |\beta_j'(x_2)\rangle \quad (11)$$

aside from an overall phase factor of no interest here [23].

The time dependence of the coherent-state amplitudes can be obtained by using the fact that

$$\hat{a}_i |\alpha_i\rangle = \alpha_i |\alpha_i\rangle \quad (12)$$

for a coherent state. The time dependence of the annihilation operators in the Heisenberg picture can be obtained as usual from their commutators with the Hamiltonian. For example

$$\frac{d\hat{a}_k}{dt} = \frac{1}{i\hbar} \left[ \hat{a}_k, \hat{H} \right] = -i\omega_k \hat{a}_k + \frac{1}{i\hbar} \sum_i h_{ki} e^{-ikx_i} \hat{b}_i. \quad (13)$$

Letting this act on the state of Eq. (11) and using Eq. (12) gives

$$\frac{d\alpha_k'(x_1)}{dt} = -i\omega_k \alpha_k'(x_1) + \frac{1}{i\hbar} \sum_i h_{ki} e^{-ikx_i} \beta_i'(x_1). \quad (14)$$

In a similar way, we also obtain

$$\begin{aligned}\frac{d\alpha_p'(x_2)}{dt} &= -i\omega_p \alpha_p'(x_2) + \frac{1}{i\hbar} \sum_j h_{pj} e^{-ipx_j} \beta_j'(x_2) \\ \frac{d\beta_i'(x_1)}{dt} &= -i\omega_0 \beta_i'(x_1) + \frac{1}{i\hbar} \sum_k h_{ki}^* e^{ikx_i} \alpha_k'(x_1) \\ \frac{d\beta_j'(x_2)}{dt} &= -i\omega_0 \beta_j'(x_2) + \frac{1}{i\hbar} \sum_p h_{pj}^* e^{ipx_j} \alpha_p'(x_2).\end{aligned} \quad (15)$$

Eqs. (14) and (15) form a set of coupled ordinary differential equations that could be solved numerically, for example. The key question is whether or not the amplitudes $\beta_i'(x_1)$ and $\beta_j'(x_2)$ that describe the state of the absorbing oscillators actually depend on the parameters $x_1$ and $x_2$. If $|\beta_i'(x_1)\rangle$ and $|\beta_j'(x_2)\rangle$ are completely independent of $x_1$ and $x_2$, then the states of the atoms can be factored out of the integral of Eq. (11) to give a product state of the form

$$|\psi(t)\rangle = \left[ \iint dx_1 dx_2 f(x_1, x_2) \prod_k |\alpha_k'(x_1)\rangle \prod_p |\alpha_p'(x_2)\rangle \right] \times \chi \prod_i |\beta_i'\rangle \prod_j |\beta_j'\rangle. \quad (16)$$

If Eq. (16) holds, then the amplitude of the entangled photon holes will have been reduced but with no decoherence between the various terms in the integrals.

In order to investigate this possibility, first consider the field and atoms in beam 1 alone for a specific value of $x_1$. This corresponds to the response of the system when a single Gaussian packet $|\alpha(x_1)\rangle$ interacts with the atomic medium. We will initially consider the response of a specific harmonic oscillator $i$ and then sum their effects. The initial state of this subsystem is given by

$$|\psi_0\rangle = \prod_k |\alpha c_k(x_1)\rangle |\beta_i(x_1,t)\rangle \qquad (17)$$

with the initial value of $\beta_i(x_1,t)$ equal to zero as before.

As in time-dependent perturbation theory, the coupling of a single atom to the field will be sufficiently small that the field will not be significantly depleted. To a first approximation, Eq. (14) reduces to

$$\frac{d\alpha_k'(x_1)}{dt} = -i\omega_k \alpha_k'(x_1). \qquad (18)$$

This equation can be solved to give

$$\alpha_k'(x_1) = \alpha c_k(x_1) e^{-i\omega_k t}. \qquad (19)$$

Here the constant $\alpha c_k(x_1)$ corresponds to the Fourier coefficients of the Gaussian pulse of Eq. (2) at the initial time $t_0 = 0$.

Inserting Eq. (19) into the second line of Eq. (15) gives the time dependence of $\beta_i(x_1,t)$ to first order in $\varepsilon$

$$\frac{d\beta_i(x_1,t)}{dt} = -i\omega_0 \beta_i(x_1,t) + \frac{1}{i\hbar}\sum_k h_{ki}^* e^{i(kx_i - \omega_k t)} \alpha c_k(x_1). \qquad (20)$$

Making use of Eq. (2) allows this to be rewritten as

$$\frac{d\beta_i(x_1,t)}{dt} = -i\omega_0 \beta_i(x_1,t) + \frac{\varepsilon \alpha c_0}{i\hbar} e^{ik_0(x_i - ct)} e^{-[x_i - (x_1 + ct)]^2 / 2\sigma_p^2}. \qquad (21)$$

Here the matrix elements $h_{ki}$ have all been assumed to be equal over the bandwidth of the pulse with a value of $\varepsilon$.

The solution to Eq. (21) becomes apparent if we factor out most of the time dependence of $\beta_i(x_1,t)$ by introducing a new variable $\zeta(t)$ defined in such a way that

$$\beta_i(x_1,t) = e^{-i\omega_0 t} \zeta(t). \qquad (22)$$

Inserting this into Eq. (21) gives

$$\frac{d\zeta(t)}{dt} = \frac{\varepsilon \alpha c_0}{i\hbar} e^{ik_0 x_i} e^{-[x_i - (x_1 + ct)]^2 / 2\sigma_p^2}. \qquad (23)$$

Integrating Eq. (23) gives

$$\zeta(t) = \frac{\varepsilon \alpha c_0}{i\hbar} e^{ik_0 x_i} \int_{-\infty}^{t} e^{-[x_i - (x_1 + ct')]^2 / 2\sigma_p^2} dt'. \qquad (24)$$

Making a change of variables to $\tau = t' - (x_i - x_1)/c$ allows this to be rewritten as

$$\zeta(t) = \frac{\varepsilon \alpha c_0}{i\hbar} e^{ik_0 x_i} \int_{-\infty}^{t-(x_i - x_1)/c} e^{-\tau^2 / 2\sigma_t^2} d\tau \\ = \frac{\varepsilon \alpha c_0}{i\hbar} \sqrt{2\pi} \sigma_t e^{ik_0 x_i} F[t - (x_i - x_1)/c]. \qquad (25)$$

Here $F(t)$ is the cumulative probability function [24] for a Gaussian distribution with standard deviation $\sigma_t = \sigma_p / c$.

The cumulative probability will rapidly approach the value $F(\infty) = 1$ after sufficient time has elapsed for the wave packet to have passed the location of the atom. Inserting the corresponding value of $\zeta(t)$ into Eq. (22) gives the final value of $\beta_i(x_1,t)$ as

$$\beta_i(x_1,t) = \frac{\varepsilon \alpha c_0}{i\hbar} \sqrt{2\pi} \sigma_t e^{ik_0 x_i} e^{-i\omega_0 t}. \qquad (26)$$

It can be seen from Eq. (26) that the final amplitude for the coherent state of harmonic oscillator $i$ is independent of the initial location $x_1$ of the Gaussian packet. Similar results can be obtained for all of the other harmonic oscillators in both beams. As a result, the final states of the harmonic oscillators all factor out of the overall system as in Eq. (16). This shows that the harmonic oscillators do not retain any "which path" information that could distinguish between the various terms in the entangled state. As a result, there is no decoherence associated with photon loss due to absorption by idealized atoms of this kind, which is one of the main results of this paper.

These results can be intuitively understood by considering an atom that is weakly driven on resonance by a continuous-wave laser at a single frequency. The probability amplitude to be in the excited state of the atom will accumulate coherently as a function of time. If we were to modulate the amplitude of the laser beam to create a Gaussian pulse (or any other shape), then the contribution of the pulse to the excited-state probability amplitude will have the same phase as that of the original laser beam regardless of when the pulse was formed by the modulator. As a result, the final state of the atom contains no information regarding the time at which the modulator created the pulse. The same conditions hold for the entangled photon hole state of Eq. (6).

The solution for $\beta_i(x_1,t)$ in Eq. (26) can be inserted back into Eq. (14) to determine the time dependence of $\alpha_k'(x_1)$ to second order in $\varepsilon$:

$$\frac{d\alpha_k'(x_1)}{dt} = -i\omega_k \alpha_k'(x_1) \\ -\alpha \frac{\varepsilon^2 \sqrt{2\pi} c_0 \sigma_t}{\hbar^2} \sum_i e^{i(k_0-k)x_i} e^{-i\omega_0 t} F[t-(x_i-x_1)/c]. \quad (27)$$

This is an iterative approach that is somewhat similar to that used in time-dependent perturbation theory. The last term in Eq. (27) will reduce the mean number of photons left in the field as required by energy conservation. It will also produce a small change in the shape of the wave packet as a result of dispersion. We are primarily interested in the coherent form of Eq. (16) and we will not need to solve Eq. (27) for the change in the amplitude of the field for reasons that will become apparent in the next section on amplification.

Eq. (16) was derived under the assumption that the changes in the field amplitudes are relatively small. For large values of $\alpha$, this can still correspond to the loss of hundreds or thousands of photons with no which-path information left in the state of the absorbing atoms. This approach could be extended to larger losses by iterating the process repeatedly, which will maintain the form of Eq. (16). That is also unnecessary if amplification is used as described in the next section.

These results show that entangled photon holes correspond to a form of decoherence-free subspace [19-21] with respect to the absorption of photons by atoms with long coherence times. All of the results in this section were based on the tacit assumption that there is negligible decay or dephasing of the excited states of the atoms or the harmonic oscillators that represent them. The effects of decoherence of the excited atomic states will be considered in Section V.

## IV. AMPLIFICATION

It was shown in the previous section that atoms with long coherence times can absorb relatively large numbers of photons without gaining any which-path information that would destroy the coherence of a large-amplitude entangled photon hole state. One might intuitively expect that the same result would hold for an amplifier aside from the noise from spontaneous emission. Amplifiers have been discussed in detail in many earlier papers [25-31] and that analysis need not be repeated here. Instead, we will focus on the question of whether or not the amplifying medium retains any which-path information that would decohere the superposition of terms in the entangled photon hole state of Eq. (6).

An amplifier can be implemented using the same two-level atoms of Fig. 4, but now all of the atoms will initially be in their excited states as illustrated in Fig. 5a. In the limit of weak interactions, the probability amplitude that a specific atom will make a transition to its ground state will be on the order of $\varepsilon^2 \ll 1$. This allows us to consider the "inverted" harmonic oscillator model illustrated in Fig. 5b as previously suggested by Glauber [27]. We label the oscillator states with an index $n$ that is 0 for the highest energy state $|0\rangle$ and increases as we go to progressively lower-energy states. Here the operator $\hat{b}^\dagger$ takes the system from state $|n\rangle$ to $|n+1\rangle$ as usual, and the only difference from an ordinary oscillator is that the energy of state $|n\rangle$ is now $-n\hbar\omega_0$ compared to that of the ground state.

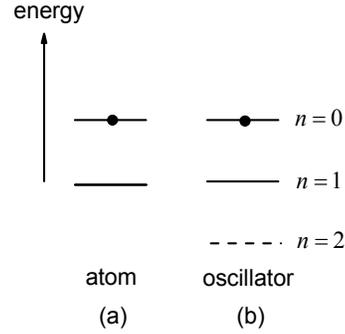

Fig. 5. (a) Inverted population of atoms used to amplify an entangled photon hole state. (b) Inverted harmonic oscillator model for the atoms in which the oscillator states are labeled in order of decreasing energy.

Energy conservation now requires that the emission of a photon be accompanied by an increase in the value of $n$ for one of the atoms. In the rotating wave approximation, the Hamiltonian of Eq. (9) becomes

$$\hat{H} = \sum_k \hbar\omega_k \hat{a}_k^\dagger \hat{a}_k + \sum_p \hbar\omega_p \hat{a}_p^\dagger \hat{a}_p - \sum_i \hbar\omega_0 \hat{b}_i^\dagger \hat{b}_i \\ - \sum_j \hbar\omega_0 \hat{b}_j^\dagger \hat{b}_j + \sum_{k,i}\left(h_{ki} e^{-ikx_i}\hat{a}_k^\dagger \hat{b}_i^\dagger + h_{ki}^* e^{ikx_i}\hat{b}_i \hat{a}_k\right) \quad (28) \\ + \sum_{p,j}\left(h_{pj} e^{-ipx_j}\hat{a}_p^\dagger \hat{b}_j^\dagger + h_{pj}^* e^{ipx_j}\hat{b}_j \hat{a}_p\right).$$

This Hamiltonian is often encountered in parametric down-conversion, squeezing, and other areas of quantum optics, where $\hat{b}_i^\dagger$ creates a photon in a second beam of light in that case.

Glauber's theorem [22] for coupled harmonic oscillators is no longer valid for the Hamiltonian of Eq. (28). Instead, we will use the fact that $\hat{a}_i^\dagger|\alpha_i\rangle \cong \alpha_i^*|\alpha_i\rangle$ for large values of $\alpha$. Even for small values of $\alpha$, we can define an operator $\hat{N}$ in such a way that

$$\hat{a}_i^\dagger |\alpha_i\rangle = \alpha_i^* |\alpha_i\rangle + \hat{N}|\alpha_i\rangle \qquad (29)$$

holds exactly. (I.e., we define $\hat{N}$ to be the difference between $\hat{a}_i^\dagger$ and $\alpha_i^*$.) The operator $\hat{N}$ can be viewed as a noise operator that reflects the effects of spontaneous emission. For example, acting on the vacuum state with $\alpha_i = 0$ gives

$$\hat{a}_i^\dagger |0\rangle = \hat{N}|0\rangle = |1\rangle. \qquad (30)$$

The expectation value of an operator $\hat{O}(t)$ in the Heisenberg picture is given by

$$\langle \hat{O}(t) \rangle = \langle \psi_0 | \hat{O}(t) | \psi_0 \rangle \qquad (31)$$

where $|\psi_0\rangle$ is the initial state of the system. As a result, we will only be interested in the value of $\hat{a}_i^\dagger(t)$ when it acts on $|\psi_0\rangle$. The fact that $|\psi_0\rangle$ is a superposition of coherent states of the field will allow the use of Eq. (29).

We begin once again by considering the situation in beam 1 for a Gaussian packet initially located at $x_1$. The time dependence of the probability amplitude to find oscillator $i$ in state $|1\rangle$ can be found from the commutator of $\hat{b}_i(x_1)$ with the Hamiltonian in analogy with Eq. (15), which gives

$$\frac{d\hat{b}_i'(x_1)}{dt} = i\omega_0 \hat{b}_i'(x_1) + \frac{1}{i\hbar}\sum_k h_{ki} e^{-ikx_i} \hat{a}_k^\dagger{}'(x_1). \qquad (32)$$

Inserting Eq. (29) into Eq. (32) gives

$$\frac{d\hat{b}_i'(x_1)}{dt} = i\omega_0 \hat{b}_i'(x_1) \\ + \frac{1}{i\hbar}\sum_k h_{ki} e^{-ikx_i} \left( \alpha_k^*{}'(x_1) + \hat{N} \right). \qquad (33)$$

Schrodinger's equation is linear, which allows us to calculate the contributions to $\hat{b}_i'(x_1)$ from the $\alpha_k^*{}'(x_1)$ terms in Eq. (33) separately from the contribution from $\hat{N}$ and then combine them later. Using the same method that was used to derive Eq. (26), we can insert the form of the wave packet from Eq. (2) into Eq. (33) to show that the contribution from the $\alpha_i^*{}'(x_1)$ terms gives

$$\hat{b}_i(x_1,t) = \frac{\varepsilon \alpha c_0}{i\hbar}\sqrt{2\pi}\sigma_p e^{-ik_0 x_i} e^{i\omega_0 t} \qquad (34)$$

after the wave packet has passed the atom.

Eq. (34) shows that the probability amplitude for an atom in the amplifier to have made a transition to its ground state is independent of the initial location of a Gaussian packet. The state of the atoms can once again be factored out to produce a product state with a form similar to that of Eq. (16). This shows that an inverted population of ideal atoms with infinite coherence times in an amplifier does not retain any which-path information that could distinguish between the superposition of terms in the entangled photon hole state.

The state of the field can be calculated using

$$\frac{d\hat{a}_k'(x_1)}{dt} = \frac{1}{i\hbar}\left[\hat{a}_k'(x_1), \hat{H}\right] \\ = -i\omega_k \hat{a}_k'(x_1) + \frac{1}{i\hbar}\sum_i h_{ki} e^{-ikx_i} \hat{b}_i^\dagger{}'(x_1). \qquad (35)$$

This requires that we first calculate the time dependence of $\hat{b}_i^\dagger{}'(x_1)$ using

$$\frac{d\hat{b}_i^\dagger{}'(x_1)}{dt} = -i\omega_0 \hat{b}_i^\dagger{}'(x_1) - \frac{1}{i\hbar}\sum_k h_{ki}^* e^{ikx_i} \hat{a}_k'(x_1). \qquad (36)$$

It is important to note the minus sign in front of the last term, which comes from the commutator $[\hat{b}_i^\dagger{}', \hat{b}_i{}'] = -1$. The solution to this equation is

$$\hat{b}_i^\dagger(x_1,t) = -\frac{\varepsilon \alpha c_0}{i\hbar}\sqrt{2\pi}\sigma_t e^{ik_0 x_i} e^{-i\omega_0 t} F[t-(x_i-x_1)/c] \qquad (37)$$

in analogy with Eq. (25). Inserting Eq. (37) into Eq. (35) and acting on $|\psi_0\rangle$ gives

$$\frac{d\alpha_k'(x_1)}{dt} = -i\omega_k \alpha_k'(x_1) \\ + \alpha \frac{\varepsilon^2 \sqrt{2\pi} c_0 \sigma_t}{\hbar^2}\sum_i e^{i(k_0-k)x_i} e^{-i\omega_0 t} F[t-(x_i-x_1)/c]. \qquad (38)$$

Eq. (38) differs from Eq. (27) by the plus sign in front of the last term. As a result, the mean number of photons increases and the field is amplified instead of attenuated. It should also be noted that the sign of the dispersion is reversed as well, which may be of practical use as will be discussed below.

These results do not include the effects of the noise operator $\hat{N}$. It is well known [25-31] that a linear (phase-insensitive) amplifier will produce spontaneous emission noise that is independent of the input state. This additive noise can reduce the signal-to-noise ratio of the system and it increases exponentially with the gain $g$ of the amplifier. Since the noise is independent of the input, its properties are the same as in earlier papers on optical amplifiers [25-31].

The effects of the amplifier noise can be reduced to a considerable extent if an amplifying medium alternates frequently with a lossy channel, as illustrated in Fig. 6. If we let the loss accumulate to the end of the channel as in Fig 6a, a relatively large gain $g$ would be required to restore the signal to its original value and the spontaneous emission noise will increase exponentially with distance. This can be mitigated by frequently amplifying the signal with a much smaller gain $g/n_A$, where $n_A$ is the number of amplifiers distributed along the channel as illustrated in Fig. 6b. In that case there is no net gain. Any noise photons from spontaneous emission will in essence not be amplified by subsequent amplifiers, since the gain is compensated by an equal amount of loss. As a result, the amplifier noise will only be proportional to the channel length rather than increasing exponentially with distance.

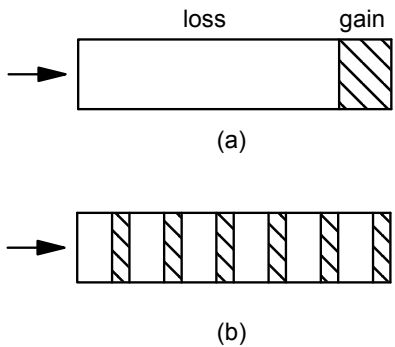

Fig.6. (a) Propagation of a beam of light through a medium with loss followed by an amplifying medium with gain $g$ that restores the field to its original amplitude. (b) Propagation through a series of media with alternating loss and individual gains of $g/n_A$, where $n_A$ is the number of amplifiers. This gives no net loss or gain, which minimizes the effects of spontaneous emission noise and dispersion.

Another benefit of using a chain of $n_A$ amplifiers is that the dispersion introduced in each section of a lossy medium can be cancelled by dispersion of the opposite sign in the adjacent amplifying medium, as can be seen by comparing Eqs. (27) and (38). This avoids distortions in the entangled photon hole state due to dispersive effects. Using alternating loss and gain also validates the theoretical approach used above in which it was assumed that the changes in the field amplitudes were small; alternating loss and gain will keep the state of the field close to its original value.

The amplifying atoms could be included as dopants such as erbium throughout the length of an optical fiber. That would result in a continuous balance between loss and gain with essentially no change in the amplitudes of the coherent states that form the basis for the entangled photon holes. The atomic coherence times must still be sufficiently long to avoid any which-path information even in that case.

Spontaneous emission noise can be avoided by using a phase-sensitive amplifier [31] and that may be a possibility for entangled photon holes. Noiseless amplification can also be achieved using post-selection [32, 33], but that appears to be limited to relatively small photon numbers.

To summarize, an amplifier consisting of an inverted population of ideal atoms (with negligible decay and dephasing) does not retain any which-path information regarding the time at which the field passed through the amplifier. As a result, such an amplifier can increase the mean number of photons by a relatively large amount without producing any significant decoherence of an entangled photon hole state, other than the usual spontaneous emission noise. Entangled photon holes can be viewed as existing in a decoherence-free subspace [19-21] with regard to amplification of this kind. The effects of spontaneous emission noise can be minimized by alternating regions of loss and amplification, and by using large values of $\alpha$ to produce a large signal-to-noise ratio even in the presence of spontaneous emission noise.

## V. ATOMIC DECOHERENCE AND BEAM SPLITTERS

The results of the two previous sections were based on the assumption that there is negligible decoherence of the atomic states over the time intervals of interest. Atomic decoherence can result from a number of sources depending on the nature of the system, including radiative decay of the excited atomic state, collisions with other atoms in a vapor, or interactions with phonons in solid-state systems. In this section, we will provide an estimate of the effects of atomic decoherence on the fidelity of the entangled photon hole states.

Nonlocal interferometry applications will typically involve two distant interferometers with unbalanced path lengths with a difference $\Delta t$ in their propagation times [1,5-14]. The output of the interferometers will depend on interference between the field amplitudes at times $t$ and $t + \Delta t$. All of the interference terms will also depend on the inner product $I$ of the atomic states that are entangled with the field amplitudes at those times, where

$$I = \prod_i \langle \psi_i(t+\Delta t) | \psi_i(t) \rangle. \tag{39}$$

Here $|\psi_i(t)\rangle$ is the state of atom $i$. Atomic decay or other decoherence mechanisms will have no effect over longer time intervals since those changes in the atomic state will be common to both of the interfering amplitudes.

It will be assumed that the initial density matrix $\rho_i(t_0)$ that describes atom $i$ in its excited state $|E_i\rangle$ will

decay exponentially in the usual way, so that the density matrix at subsequent times has the form

$$\rho_i(t) = e^{-t/\tau_D} |E_i\rangle\langle E_i| + \rho_\perp(t). \quad (40)$$

Here $\rho_\perp(t)$ corresponds to a mixture of states that are orthogonal to $|E_i\rangle$ and therefore do not contribute to the inner product $I$. The amplitude $A_I$ of the interference terms will be reduced to

$$A_I = c_I e^{-n_L \Delta t / \tau_D}. \quad (41)$$

Here $c_I$ is a constant that depends on the details of the interferometer and $n_L$ is the number of atoms that were initially left in the excited state; this is also the number of photons that were lost due to absorption in the medium.

It is apparent from Eq. (41) that the visibility of the nonlocal interference will depend on the ratio $R = \Delta t / \tau_D$. Atomic vapors typically have coherence times ranging from 10 to 300 ns. It should be feasible to use interferometers where $\Delta t$ is a small fraction of a nanosecond. Under those conditions, the number of photons that can be lost to absorption while maintaining a high visibility is given by

$$n_L << \frac{1}{R} \sim 10^3. \quad (42)$$

Thus a relatively large number of photons can be lost due to absorption provided that R is sufficiently small. Similar results apply to the case of amplification.

The actual value of the atomic coherence time will depend strongly on the nature of the medium. Transmission of quantum information over large distances in optical fibers is of particular interest. Impurities and other imperfections in optical fibers may eventually be reduced to the point that the residual loss is primarily due to coherent effects such as Brillouin or Raman scattering. The corresponding coherence time $\tau_D$ would be that of the scattered photons and phonons in the case of Brillouin scattering. This is a coherent process that involves phase matching and energy conservation, and it may be that $\tau_D$ would be very long if the fiber were cooled to low temperatures to reduce the effects of thermal phonons, for example. Those issues are beyond the intended scope of this paper and further research in that area would be desirable.

Decoherence of entangled photon holes can also occur if a beam splitter is placed in the transmission channel. Consider the effect that this would have on two terms in the entangled state of Eq. (6) that correspond to probability amplitudes for Gaussian packets initially located at $x_1$ and $x_1 + c\Delta t$. They will produce weak fields in the output port of the beam splitter that are distinguishable at least in principle. There will now be an inner product analogous to Eq. (39) but involving the fields in the output port instead. If the width of the packets is much smaller than $c\Delta t$, then their inner product will be limited to the vacuum state components of each. This gives a visibility for the interference pattern that is proportional to $\exp[-\alpha_S^* \alpha_S] = \exp[-\bar{n}_S]$, where $\alpha_S$ is the amplitude for the coherent state in the output port and $\bar{n}_S$ is the corresponding mean number of photons. This shows that the visibility of the interference pattern would be substantially reduced if even a single photon were removed from the beams using a beam splitter.

The strong loss of visibility due to the presence of a beam splitter can be beneficial in the sense that it shows that an eavesdropper cannot simply split off a "copy" of the information in a quantum key distribution system based on entangled photon holes. On the other hand, it precludes the use of pre-existing optical fiber networks that typically contain relatively poor connectors, switches, etc. Dedicated optical fibers or free-space links would be required instead, and they could have relatively small beam splitter losses. Even then, the overall amplitude $\alpha$ would have to be limited to moderate values. This may result in a trade-off between using small values of $\alpha$ to limit the effects of beam splitter losses versus larger values of $\alpha$ to improve the signal to noise ratio.

## VI. SUMMARY AND CONCLUSIONS

It has been shown that relatively large numbers of photons can be absorbed from an entangled photon hole state with no adverse effects if the atoms in the medium have no significant decay or dephasing over the time interval $\Delta t$ that is characteristic of nonlocal interferometers [5-14]. Roughly speaking, the atoms that are excited as a result of absorbing a photon do not retain any which-path information that can be used to determine when a photon hole passed through the medium. The same is true for the amplification of entangled photon holes using an inverted population of atoms. As a result, entangled photon holes can be viewed as occupying a decoherence-free subspace [19-21] with respect to interactions with ideal atoms.

Entangled photon holes continue to show a reduced sensitivity to loss and amplification in the more realistic situation in which the excited atomic states decohere with a time constant $\tau_D$. In that case the number of photons that can be absorbed without appreciable degradation is $n_L \sim \tau_D / \Delta t$. This ratio depends a great deal on the nature of the lossy medium, and it can have relatively large values for an atomic vapor, for example. Further research on possible ways to achieve large values of $\tau_D$ in optical fibers for use in quantum communication systems would be desirable.

Entangled photon holes are very sensitive to loss due to the insertion of a beam splitter, and dedicated optical fibers or free-space links would be required as a result. The fact that entangled photon holes have low sensitivity to photon loss by atomic absorption but high sensitivity to beam splitter losses suggests that the effects of photon loss cannot always be adequately represented by a series of beam splitters, as is often done in practice [34, 35]. This also implies that there are limitations on the use of density matrix techniques in combination with the Markov approximation [36], in which it is assumed that the environment rapidly decoheres. If the absorbing atoms were considered to be part of the "environment", then the use of the Markov approximation would give totally different results.

It is probably apparent that the reduced sensitivity to loss and amplification is not specific to this particular form of entangled state. Methods for generating entangled photon holes with large values of $\alpha$ and for implementing nonlocal interferometry will be discussed in a subsequent paper, along with the possibility of reduced sensitivity to beam splitter losses. Those topics are beyond the intended scope of this paper, which is already lengthy.

These results are of fundamental scientific interest. Whether or not they are of practical use in quantum communications or quantum key distribution will require further theoretical and experimental work, including an investigation of the properties of coherent loss mechanisms in optical fibers.

**ACKNOWLEDGEMENTS**

I would like to acknowledge stimulating discussions with Brian Kirby and Todd Pittman. This work was supported by grant # W31P4Q-10-1-0018 from the DARPA Defense Sciences Office (DSO).